\documentclass[prl,twocolumn,showpacs,amssymb,floatfix]{revtex4}
\usepackage{graphicx}
\newcommand{\be}{\begin{eqnarray}} \newcommand{\ee}{\end{eqnarray}}

\begin{document}

\title{Fronts with a Growth Cutoff but Speed Higher than $v^*$}
\author{Debabrata Panja}
\affiliation{ Instituut--Lorentz, Universiteit Leiden, Postbus 9506, 2300 RA
Leiden, The Netherlands}
\author{Wim van Saarloos}
\affiliation{ Instituut--Lorentz, Universiteit Leiden, Postbus 9506, 2300 RA
Leiden, The Netherlands}

\date{\today}

\begin{abstract} 
Fronts, propagating into an unstable state $\phi=0$, whose asymptotic
speed $v_{\text{as}}$ is equal to the linear spreading speed $v^*$ of
infinitesimal perturbations about that state (so-called pulled fronts)
are very sensitive to changes in the growth rate $f(\phi )$ for $\phi
\ll 1$. It was recently found that with a small cutoff, $f(\phi)=0$
for $\phi < \varepsilon$, $v_{\text{as}}$ converges  to $v^*$ very
slowly from below, as $\ln^{-2} \varepsilon$. Here we show that with
such a cutoff {\em and} a small enhancement of the growth rate for
small $\phi$ behind it, one can have $v_{\text{as}} > v^*$, {\em even}
in the limit $\varepsilon \to 0$. The effect is confirmed in a
stochastic lattice model simulation where the growth rules for  a few
particles per site are accordingly modified.

\end{abstract}


\maketitle

Pulled fronts are those fronts that propagate into a linearly unstable
state, and whose  asymptotic front speed $v_{\text{as}}$ equals  the
linear spreading  speed $v^*$ of infinitesimal perturbations about the
unstable state \cite{bj,vs2,ebert}. The name pulled front refers to
the picture that in the leading edge of these fronts, the perturbation
about the unstable state grows and spreads with speed $v^*$, while the
rest of the front gets ``pulled along'' by the leading edge. That this
notion is not merely an intuitive picture but can be turned into a
mathematically precise analysis is illustrated by the recent
derivation of exact results for the general  power law convergence of
the front speed to the asymptotic value $v^*$ \cite{ebert}. Fronts
which propagate into a linearly unstable state and whose asymptotic
speed $v_{\text{as}} >v^*$ are refered to as pushed, as it is the
nonlinear growth in the region behind the leading edge that pushes
their front speed to higher values.  If the state is not linearly
unstable, then $v^*$ is trivially zero; in such cases the front
propagation is always dominated by the nonlinear growth in the front
region itself, and hence fronts in this case are in a sense ``pushed''
too.

For the field $\phi(x,t)$, the dynamics of  fronts that we
consider in this paper is given by the usual  nonlinear diffusion equation
\begin{eqnarray}
\frac{\partial\phi}{\partial t}\,=\,\frac{\partial^2\phi}{\partial
x^2}\,+\,f(\phi)\,.
\label{e1}
\end{eqnarray}
In the standard case,  the growth function $f(\phi)$ has the form
$f(\phi)=\phi-\phi^n$, with $n>1$. Equation (\ref{e1}) has two
stationary states for $\phi(x,t)$: $\phi(x,t)=0$ and $\phi(x,t)=1$. Of
these, $\phi(x,t)=1$ is stable and $\phi(x,t)=0$ is unstable. The
asymptotic speed of (pulled) fronts propagating from $\phi(x,t)=1$
into $\phi(x,t)=0$ in Eq. (\ref{e1}) is $v^*=2$.

\begin{figure}[tb]
\begin{center}
\includegraphics[width=0.98\linewidth]{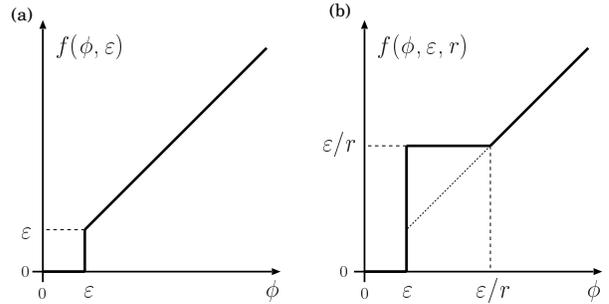}
\end{center}
\vspace*{-2mm}
\caption{ {\em (a)} Shape of the function $f(\phi,\varepsilon)$ used
by Brunet and Derrida to study the effect of a finite particle cutoff
in the growth rate on the front speed.  {\em (b)} The growth function
$f(\phi,\varepsilon,r)$ (thick line) we analyze in this paper. In both
cases we have kept only the linear term 
of $f(\phi)$ to plot the graphs, since $\varepsilon\ll1$, so that  the
nonlinear terms in $f$ are much smaller than the
linear terms. } \label{fig1}
\end{figure}
The sensitivity of pulled fronts to the precise dynamics for small
perturbations about the unstable state has recently surfaced in a
remarkable way \cite{bd}.  Often, in equations like (\ref{e1}), the
field $\phi(x,t)$  is the density of particles in a continuum
description. If one then considers fronts in stochastic particle model
versions of (\ref{e1}), the linear growth term in $f(\phi)$ implies
that for small particle density, the rate at which new particles are
created is proportional to the density itself. Brunet and Derrida
\cite{bd} were the first to realize  the fact that for new particles
to be created in any given realization, the density must be at least
one ``quantum'' of particle density strong, and that this provides a
natural lower cutoff for the growth that strongly affects the front
speed. Indeed, to mimic  this effect,  they considered a deterministic
front of the type in Eq. (\ref{e1}) with $n=3$, and by hand introduced
a cutoff  of the type sketched in Fig.~\ref{fig1}a in the growth
function at $\phi=\varepsilon \ll 1$. In this paper, we denote their
growth function by $f(\phi,\varepsilon)\equiv
[\phi-\phi^3]\,\Theta(\phi-\varepsilon)$ where $\Theta$ is the unit
step function.  For small $\varepsilon$, the asymptotic front speed
$v_{\text{as}}(\varepsilon)$ was then found to be \cite{bd}
\begin{eqnarray}
v_{\text{as}}(\varepsilon) \, \simeq
\,v^*\,-\,\frac{\pi^2}{(\ln\varepsilon)^2} + \cdots\,.
\label{e2}
\end{eqnarray}
Brunet and Derrida subsequently identified  $\varepsilon$ with  $1/N$,
where $N$ is the average number of particles at the saturation state
of the front, corresponding to the stable state $\phi(x,t)=1$ of the
density field. The slow logarithmic convergence to the asymptotic
front speed from below as a function of $N$, implied by
Eq. (\ref{e2}), has been confirmed in various studies of stochastic
lattice models \cite{bd,breuer,vanzon,kns,levine,PvS}. Note that  for
$\phi < \varepsilon$, the growth function $f$ vanishes, and as a
result, strictly speaking, the state $\phi=0$ is not linearly
unstable; hence fronts in this model are always weakly pushed for any
nonzero value of $\varepsilon$ \cite{deb2}.

In this paper, we demonstrate an even more surprising aspect of the
sensitivity to small changes in the growth function $f$ of the
``pulled'' fronts, that we have at $\varepsilon=0$: if $f$ is
sufficiently {\em enhanced} in a range of $\phi$ of the order of
$\varepsilon$, the asymptotic front speed $v_{\text{as}}$ can become
larger than $v^*$ and {\em not converge to $v^*$ as} $\varepsilon \to
0$. For fluctuating fronts, this implies that if the stochastic growth
rates for small occupation densities $n_i$ are somewhwat enhanced over
a linear behavior $\sim n_i$, then such stochastic fronts may move
faster than $v^*$ and {\em never converge to their naive mean field
limit for}  $N\to \infty$. This effect may be of relevance
for the coarse-grained field theory for DLA, as it is empirically
known to be essential to modify the growth function for small cluster
densities \cite{brener}.

We now discuss our results first, and then summarize their
derivation.

To be specific, we  consider the nonlinear diffusion equation
(\ref{e1}) with the growth function sketched in Fig.~\ref{fig1}b,
\begin{eqnarray}
f(\phi,\varepsilon,r)\,&=&
\,f(\phi,\varepsilon)\quad\mbox{for}\,\,\phi<\varepsilon\,\,\,\mbox{and
for}\,\,\,\phi>\varepsilon/r\nonumber\\&=
&\,\varepsilon/r\quad\quad\,\,\,\mbox{for}\,\,\,\varepsilon\leq\phi\leq\varepsilon/r\,,
\label{e3}
\end{eqnarray} 
with  $r<1$.  We show that while for any fixed value of $r$,
\begin{eqnarray}
\lim_{\varepsilon\rightarrow0}f(\phi,\varepsilon,r)\,=\,f(\phi)\,,
\label{e4}
\end{eqnarray} 
the asymptotic front speed $v_{\text{as}}(\varepsilon,r)$ has the
property that
\begin{eqnarray}
\lim_{\varepsilon\rightarrow0}\,v_{\text{as}}(\varepsilon,r)&=&v^*\quad\quad\mbox{for}\,\,\,r>r_c\,,\nonumber\\
\lim_{\varepsilon\rightarrow0}\,v_{\text{as}}(\varepsilon,r)&> &
v^*\quad\quad\mbox{for}\,\,\,r<r_c\,.
\label{e5}
\end{eqnarray} 
where
\begin{eqnarray}
r_c\,=\,\frac{1\,+\,e^{-\,(v^{*2}\,-\,2)}}{v^{*2}}\,=\,0.283833\ldots\,.
\label{erc}
\end{eqnarray} 
Hereafter, to save writing, we denote $v_{\text{as}}(\varepsilon,r)$
simply by $v$. For $\varepsilon\rightarrow0$, the asymptotic speed at
a given value of $r\leq r_c$ in our model is given by the relation
\begin{eqnarray}
r=\frac{1}{v^2}\left[1+\frac{\displaystyle{1-\frac{2}{v}\sqrt{\frac{v^2}{4}-1}}}{\displaystyle{1+\frac{2}{v}\sqrt{\frac{v^2}{4}-1}}}\,\,e^{-\left\{v^2-\frac{2}{{1+\frac{2}{v}\sqrt{\frac{v^2}{4}-1}}}\right\}}\right],
\label{e6}
\end{eqnarray} 
from which the value of $r_c$, given by Eq. (\ref{erc}), follows.
\begin{figure}[tbf]
\begin{center}
\includegraphics[width=0.44\textwidth]{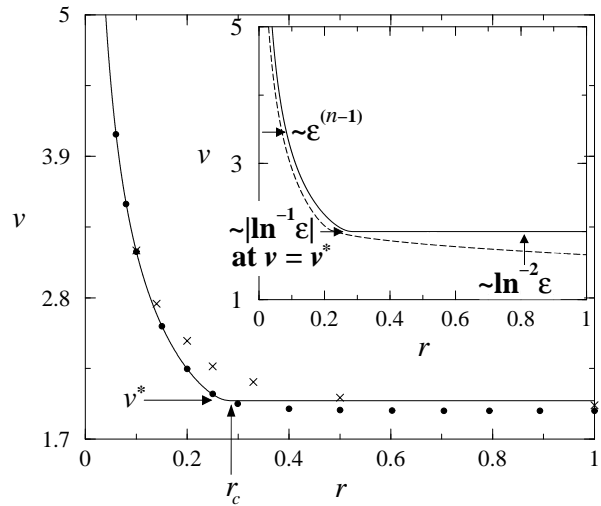}
\end{center}
\vspace*{-3mm}
\caption{Comparison of simulation data for
$v_{\text{as}}(\varepsilon,r)$ with the analytical prediction
(\ref{e6}), which is plotted as the solid line. The solid dots in
Fig. \ref{fig2} represent the numerical data for Eq.~(\ref{e3}) with
$\varepsilon=2\times10^{-5}$ and $n=3$. The crosses are the data
points for fronts in the stochastic growth model described in the
text. Inset: illustration of the leading order rate of
convergence of the $v_{\text{as}}(\varepsilon,r)$  curve to the
$\varepsilon\rightarrow 0$ limit, by means of the schematic dashed 
curve. \label{fig2}}
\end{figure}
\begin{figure}[tbf]
\begin{center}
\includegraphics[width=0.33\textwidth,angle=270]{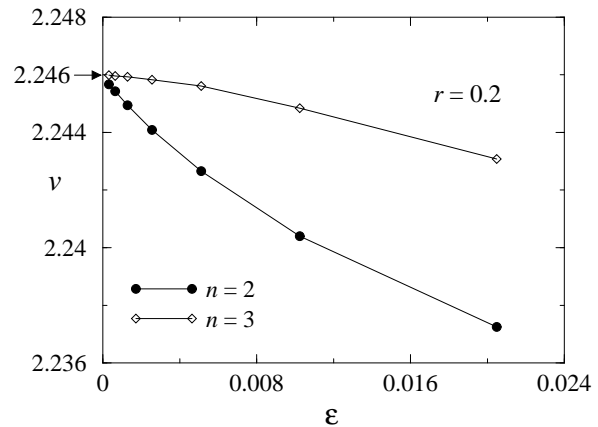}
\end{center}
\vspace*{-4mm}
\caption{Numerical data for  $v_{\text{as}}(\varepsilon,r)$ as a
function of $\varepsilon$ for $n=2$ and $3$, at $r=0.2$. The graph
demonstrates the insensitivity of $v$ to $\varepsilon$ for small
values of $\varepsilon$ (note the fine scale on the vertical axis),
as well as the convergence as $\varepsilon^{n-1}$  to its
$\varepsilon\rightarrow0$ value $\simeq2.246$, given by
Eq. (\ref{e6}), for two different values of $n$.
\label{fig3}}
\end{figure}

These expressions show that the limits do not commute for $r<r_c$:
taking the limit $\varepsilon \to 0$ first in $f$ yields a front speed
$v^*$ but the limit $v_{\text{as}}(\varepsilon \to0, r) > v^*$.  The
reason is that for $r<r_c$ there is always a little tail of the front
that runs faster than $v^*$ and makes $\phi$ nonzero. Once $\phi$ is
nonzero, growth continues and the region behind it just has to follow
it with the same asymptotic speed.

Our analysis is corroborated by  numerical results obtained by solving
Eq. (\ref{e1}) forward in time, (with Gaussian initial conditions).
The data for $v$ vs. $r$ at $\varepsilon=2\times10^{-5}$ are shown as
solid dots in Fig. \ref{fig2}.  Note that for $r<r_c$, the solid dots
fall on top of our prediction (\ref{e6}) drawn with a solid line,
while for $r>r_c$, they systematically fall below the solid line
$v=v^*$. The reason for it is the difference between the rates of
convergence as $\varepsilon\rightarrow0$, which is illustrated
in the  inset of Fig.~\ref{fig2} by means of the schematically drawn 
dashed line. The arrows in the inset indicate the rate of convergence 
of the dashed $r$-$v$ curve towards the limiting one, given by Eq. 
(\ref{e6}). For $r>r_c$, the convergence is $\sim \ln^{-2}\varepsilon$ 
like in the case for $r=1$, analyzed in Ref. \cite{bd}; but for 
$r<r_c$ the convergence is much faster, $\varepsilon^{n-1}$. This 
latter behavior is illustrated for $r=0.2$ and $n=2,3$ in 
Fig.~\ref{fig3} --- note the fine scale on the vertical axis!

That the effect of increasing asymptotic speed with decreasing $r$
below $r_c$ is a {\it real\/}  effect for stochastic fronts too is
illustrated by the crosses in Fig.~\ref{fig2}: these represent the
data for the average speed of fronts in a  reaction-diffusion system
X$ \leftrightharpoons 2$X, for discrete X particles on a lattice with
$N=10^4$ \cite{correspond}, where the growth rates  have been modified
when the number of particles $n_i$ on a lattice site $i$ is less than
$1/r$. In accord with the shape of the growth function $f$ illustrated
in Fig.~\ref{fig1}b, the rate at which particles are created at a
lattice site $i$ with $1 \le n_i < 1/r$ particles is simply taken to
be the same as the rate for $n_i=1/r$ (corresponding to the integral
values $1/r=1,2,3,4,5,7$ and $10$, due  to the discreteness of
particles). As one can see from Fig. \ref{fig2}, already when $r=0.5$,
i.e., when only the growth rate at lattice sites with one particle is
increased by a factor 2, the asymptotic growth speed is above the
value $v^*=2$.

In the remainder of this paper, we derive the analytical results for
the nonlinear diffusion equation with the growth function (\ref{e3}).
Our analysis is based on the following observation: for
$\varepsilon=0$, it is well known that the nonlinear diffusion
equation allows a continuous family of front solutions with $v\ge
v^*$. When  such fronts solutions are parametrized by their velocity
$v$, and when the growth rate is modified to allow a transition to a
``pushed'' front with velocity $v^\dagger$, it is also known
\cite{vs2,ebert} that solutions with $v< v^\dagger$ are unstable to a
localized mode. In our analysis, we therefore consider a front with a
given fixed velocity $v$ and, for small $\varepsilon$, determine when
upon decreasing $r$  a localized mode of the stability operator
crosses the eigenvalue zero. In the limit $\varepsilon \to 0$ this
marks the selected pushed front in the $r$-$v$ diagram.

\begin{figure}[tb]
\begin{center}
\includegraphics[width=0.44\textwidth]{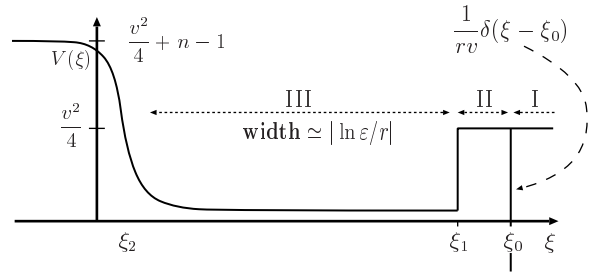}
\end{center}
\vspace*{-3mm}
\caption{The potential $V(\xi)$ for $v\geq v^*$ and infinitesimally small
$\varepsilon$ in the Schr\"odinger operator that determines the
temporal eigenvalues of  the
stability analysis. $\xi_2$ marks the position of the region of finite
width where the potential crosses over from the asymptotic value on
the left where $\phi_{\text{as}} \approx 1$ to the  value in the well where
$\phi_{\text{as}} \ll 1$, $\xi_1$ the position of the step and $\xi_0$
the position of the delta function term in the potential. }
\label{fig4}   
\end{figure}

To carry out the linear stability analysis of the front solution, it
is convenient to follow the standard route of transforming the linear
eigenvalue equation into a Schr\"odinger eigenvalue problem
\cite{bj,ebert}. We consider a function $\phi(x,t)$, which is
infinitesimally different from the asymptotic front solution
$\phi_{\text{as}}(\xi)$ in the comoving frame $\xi=x-vt$, i.e.,
$\phi(x,t)=\phi_{\text{as}}(\xi)+\eta(\xi,t)$.  Upon linearizing
Eq. (\ref{e1}) in the comoving frame, one finds that the function
$\eta(\xi,t)$ obeys the following equation:
\begin{eqnarray}
\frac{\partial\eta}{\partial
t}\,=\,v\,\frac{\partial\eta}{\partial\xi}\,+\,\frac{\partial^2\eta}{\partial\xi^2}\,+\,\frac{\delta
f(\phi)}{\delta\phi}\bigg|_{\phi\,=\,\phi_{\text{as}}}\,\eta\,.
\label{e13}
\end{eqnarray}
Since this equation is linear in $\eta$, the question of stability can
be answered by studying the spectrum of the temporal eigenvalues. To
this end, we express $\eta(\xi,t)$ as
\begin{eqnarray}
\eta(\xi,t)\,=\,e^{-\,Et}\,e^{-\,v\xi/2}\,\psi_E(\xi)\,,
\label{e14}
\end{eqnarray}
which converts Eq.~(\ref{e13}) to a one-dimensional Schr\"odinger
equation  for a particle in a potential with $\hbar^2/2m=1$:
\begin{eqnarray}
\left[\,-\,\frac{d^2}{d\xi^2}+\frac{v^2}{4}-\frac{\delta
f(\phi)}{\delta\phi}\bigg|_{\phi\,=\,\phi_{\text{as}}}\right]\psi_E(\xi)=\,E\,\psi_E(\xi),
\label{e15}
\end{eqnarray}
In Eq.~(\ref{e15}), the quantity
$\displaystyle{V(\xi)=\left[\,\frac{v^2}{4}\,-\,\frac{\delta
f(\phi)}{\delta\phi}\bigg|_{\phi\,=\,\phi_{\text{as}}}\right]}$ plays
the role of the potential. It is easily obtained explicitly from the
expression (\ref{e3}) for $f(\phi,\varepsilon,r)$ as
\begin{eqnarray}
V(\xi)&=&
\left[\,\frac{v^2}{4}-1+n\phi_{\text{as}}^{n-1}(\xi)\right]\Theta(\xi_1-\xi)
\nonumber
\\  & & ~~~~ + \frac{v^2}{4}\,\Theta(\xi-\xi_1)-
\frac{1}{rv}\,\delta(\xi-\xi_0)\, , 
\label{e16}
\end{eqnarray}
where $\phi(\xi_0)=\varepsilon$ and $\phi(\xi_1)=\varepsilon/r$.  The
form of the potential for $v>v^*$ and  small $\varepsilon$ is sketched
in Fig.~\ref{fig4}. Keep in mind that $\phi_{\text{as}}(\xi)$ is a
monotonically increasing function from $\varepsilon/r$ at $\xi_1$ {\it
towards the left\/}, and that $\phi_{\text{as}}(\xi \to -\infty)=1$.
As a result, in Fig. \ref{fig4}, $V(\xi)$ also increases monotonically
towards the left for $\xi<\xi_1$. On the right of $\xi_1$, $V(\xi)$ is
constant at $v^2/4$, and at $\xi_0$, there is an attractive
$\delta$-function potential of strength $(rv)^{-1}$ \cite{notedelta}.
The crucial feature for the stability analysis below is  the fact that
$V(\xi)$ stays remarkably flat at a value $2\varepsilon/r$ over a
distance $(\xi_1-\xi_2)\simeq|\ln\varepsilon/r|$ \cite{deb2}, and on
the left of $\xi_2$, it increases to the value of $v^2/4+n-1$, over a
distance of order unity.

If there exist negative eigenvalues of the Schr\"odinger equation
(\ref{e15}), then according to Eq. (\ref{e14}), $\eta(\xi,t)$  grows
in time in the comoving frame, i.e., the front solution
$\phi_{\text{as}}(\xi)$ is unstable. For our purpose, therefore, we
look for the value of $r$ at which there is a bound state  of
Eq.~(\ref{e15}) with eigenvalue $E$, such that $E\rightarrow0-$  for
the potential sketched in Fig.~\ref{fig4}. This is a problem in
elementary quantum mechanics. For $\varepsilon\rightarrow0$, the
potential $V(\xi)$ is essentially constant in the  left neighbourhood
of $\xi_1$, and hence for $v >v^*$ and $E\rightarrow0-$, $\psi_E(\xi)$
can be written as
\begin{eqnarray}
\psi_E(\xi)\,&=&
\,A_{-}e^{\lambda_1(\xi-\xi_1)}\quad\quad\quad\quad\quad\quad~\mbox{for}\,\,\xi\leq\xi_1\nonumber\\
&=&
\,A\,e^{\lambda_2(\xi-\xi_0)}+B\,e^{\lambda_2(\xi_0-\xi)}\quad\mbox{for}\,\,\xi_1
\leq \xi\leq\xi_0\nonumber\\&=
&\,A_2\,e^{-\lambda_2(\xi-\xi_0)}\quad\quad\quad\quad\!\quad\quad\mbox{for}\,\,\xi>\xi_0\,,
\label{e17}
\end{eqnarray}
where $\lambda_1=\sqrt{v^2/4-1}$ and $\lambda_2=v/2$. The function
$\psi_E(\xi)$ must be continuous at $\xi_1$ and  $\xi_0$, while its
slope is continuous at $\xi_1$, but not at $\xi_0$. Matching of these
boundary conditions to determine the value of $r$, where the bound
state eigenvalue $E$ crosses zero, also requires an expression for the
distance $\xi_0-\xi_1$. To this end, we divide the range of $\phi$
values between $0$ and $1$ into the three regions marked in
Fig. \ref{fig4}: {\it (i)\/} region I, where
$\phi_{\text{as}}<\varepsilon$, {\it (ii)\/} region II, where
$\varepsilon\leq\phi_{\text{as}}<\varepsilon/r$, and {\it (iii)\/}
region III, where $\phi_{\text{as}} \geq\varepsilon/r$. In the
comoving frame, the asymptotic shape $\phi_{\text{as}}(\xi)$ of the
front is  the solution of the  differential equation $\phi^{\prime
\prime}_{\text{as}}\,+\,v\,
\phi^{\prime}_{\text{as}}\,+\,f(\phi_{\text{as}},\varepsilon,r)\,=\,0\,$,
where a prime denotes a derivative with respect to $\xi$.  The
solutions of $\phi_{\text{as}}(\xi)$ in the regions I and II that
satisfy the continuity of $\phi_{\text{as}}(\xi)$ and
$\phi_{\text{as}}'(\xi)$, are respectively given by
\begin{eqnarray}
\phi_{\text{as}}(\xi)\,=\,\varepsilon\,e^{-\,v(\xi\,-\,\xi_0)}\quad\quad\quad\mbox{and}\nonumber\\&&\hspace{-5.2cm}\phi_{\text{as}}(\xi)=\left[\varepsilon-\frac{\varepsilon}{rv^2}\right]e^{v(\xi_0-\xi)}+\frac{\varepsilon\,(\xi_0-\xi)}{rv}+\frac{\varepsilon}{rv^2}.
\label{e4c}
\end{eqnarray}
The length $\xi_0-\xi_1$ of region II is obtained by equating
$\phi_{\text{as}}(\xi_1)$ from the second line of Eq.~(\ref{e4c}) to
$\varepsilon/r$. After dividing out a factor of $\varepsilon/r$, this
condition becomes
\begin{eqnarray}
\left[\,r\,-\,\frac{1}{v^2}\right]\,e^{v(\xi_0-\xi_1)
}+\,\frac{\xi_0-\xi_1 }{v}\, +\,\frac{1}{v^2}\, =\,1\,.
\label{e8}
\end{eqnarray}
Thereafter, using Eqs.~(\ref{e17}) and (\ref{e4c}), one arrives at
Eq.~(\ref{e6}). 

The above analysis yields the relation between $v$ and the critical
value of $r$ in the limit $\varepsilon \to 0$. The convergence with
$\varepsilon$, i.e., the rate of  approach with $\varepsilon$ of the
dashed curve to the solid one in Fig.~\ref{fig2},  can be obtained by
considering the effect of the  $n\phi_{\text{as}}^{n-1}(\xi)$ term of
$V(\xi)$  on the eigenfunctions and eigenvalues.  For $v> v^*$,  this
term is simply a correction of order $\varepsilon^{n-1} $ to the
finite bottom value of the potential. This term can be included
perturbatively, and accordingly it leads to a shift of order
$\varepsilon^{n-1}$ in the critical value of $r$. As Fig.~\ref{fig3}
illustrates, this prediction  is confirmed numerically. The case
$v=v^*$ calls for a more detailed analysis, since the bottom value of
the potential vanishes in the limit $\varepsilon \to 0$. In this case,
it is known \cite{ebert,bd} that
$\phi_{\text{as}}(\xi)\sim(C\xi+D)e^{-\xi}$, so $V(\xi)\simeq n
(\varepsilon /r) ^{n-1} (\xi/\xi_1)^{n-1} e^{(n-1)(\xi_1-\xi_)}$ in
the leading order of $\varepsilon$. In dominant order, we need to keep
only the exponential behaviour, and the solution of $\psi_E(\xi)$ is
then given by the Bessel function $A_-
K_0(2\sqrt{n\varepsilon^{n-1}e^{-(n-1)(\xi_1-\xi)}/r^{n-1}}\,)$ in the
left neighbourhood of $\xi_1$. The $|\ln^{-1}\varepsilon|$ scaling for
the asymptotic approach of the dashed curve to the solid one is then
easily obtained once the boundary conditions at $\xi_1$ and $\xi_0$
are matched with the use of Eq.~(\ref{e8}).

The logarithmic convergence of $v$ to $v^*$ from below for $r>r_c$ can
be understood from  an argument along the lines of that for $r=1 $
\cite{bd}. For $v< v^*$, the front profile $\phi_{\text{as}}(\xi)$ in
region III is of the form $\phi_{\text{as}}(\xi)\sim C \sin[k
(\xi-\xi_2) + \beta] e^{-\xi}$. For $r=1$, region II is absent; in
that case, the matching to the profile in region  I and the divergence
of the width $\xi_0-\xi_2 \simeq |\ln \varepsilon |$ implies $k \simeq
\pi|\ln \varepsilon|^{-1}$. For $r_c < r<1$, the matching to region II
will change the prefactor, but $k$ will still scale as $|\ln
\varepsilon|^{-1}$ because the width of region III still diverges
logarithmically.  As for $r_c\leq r<1$, this translates into a scaling
of $v^*-v$ as $| \ln \varepsilon |^{-2}$, with a prefactor that
depends on $r$.  Note that this scaling is nicely consistent with the
convergence of the $r$-$v$ curve towards the point $(r_c,v^*)$ from
the left, due to the fact that the slope of this curve vanishes at
this point, and the convergence from below to this point scales  as
the square of the convergence from the left.

We finally end this paper with the note that if the (nonnegative)
growth rate is bounded from above by $f(\phi,\varepsilon)$ in the
interval $\phi\leq\varepsilon/r$, but is equal to
$f(\phi,\varepsilon)$ for $\phi>\varepsilon/r$, then as
$\varepsilon\rightarrow0$, the asymptotic front speed  converges to
$v^*$ with the {\it same\/} logarithmic convergence of Eq.~(\ref{e2})
for {\it any\/} $r$. It simply follows from the inequality
$v_{\text{as}}(\varepsilon/r)<v<v_{\text{as}}(\varepsilon)$, where
$v_{\text{as}}(\varepsilon)$ is given by Eq. (\ref{e2}).

D. P. wishes to acknowledge financial support from ``Fundamenteel 
Onderzoek der Materie'' (FOM).


\begin{thebibliography}{02}
\bibitem{notedelta} The $\delta$-function in Eq.~(\ref{e16}) appears
from the functional derivative of $f$ in $V$, since there is a
discontinuity of magnitude $\varepsilon$ in
$f(\phi_{\text{as}},\varepsilon,r)$ at
$\phi_{\text{as}}=\varepsilon$. This discontinuity contributes an
amount equal to ${\varepsilon}{r^{-1}}\, {d \Theta
(\phi_{\text{as}}-\varepsilon)}/ {d\phi_{\text{as}}}\,=\,
{\varepsilon}{[r\,|\phi_{\text{as}}'(\xi_0)|]^{-1}}\,\delta(\xi-\xi_0)$
to $V(\xi)$. As $\phi^{\prime}_{\text{as}}= \varepsilon r$ according
to Eq.~(\ref{e4c}),  the prefactor $1/(rv)$.
\end{thebibliography}
\end{document}